%% file: E7Aug7-8Final.tex
\font\eightrm=cmr8 at 8pt
\documentclass{oxarticle}
\usepackage{amsmath, amssymb}
\usepackage{graphics, setspace}
\usepackage{epstopdf}
\usepackage{bm}
\usepackage{tikz}
\usepackage{xcolor}
 \input{Johnmacros29}

\newcounter{orange} 
\newcounter{apple} 
\addtocounter{apple}{1}
\newcounter{grape} 
\addtocounter{grape}{1}
\newcommand{\numberhere}{TheEPapersAug1\;}
\newcommand{\articlenumber}{E7Aug7-8Final} 

\usepackage{color}

\newcommand{\mathsym}[1]{{}}
\newcommand{\unicode}[1]{{}}

\usepackage{tikz}

\begin{document}
%\pagestyle{empty}

%\vspace*{1in}
 \begin{center}
{   \huge The Free Massive Quadratic Action\\ from the Exotic Model (E7) 
\\
[1cm]}  
%{    J. A. Dixon}
%

\renewcommand{\thefootnote}{\fnsymbol{footnote}}
%\footnotetext[1]{~here we have a footnote.}\renewcommand{\thefootnote}\arabicfootnote}}

{ John A. Dixon\footnote{jadixg@gmail.com, john.dixon@ucalgary.ca}\\Physics Dept\\University of Calgary 
\\[1cm]}  
%\vfill
 \end{center}
 
 \Large

 \begin{center} Abstract
 \end{center}

The Exotic Model (``the XM"),   arises from  adding a special exotic invariant to the \SSM.   The XM has a \susy\ violating mass spectrum without using spontaneous or explicit breaking of \susy\ (``SUSY").   The splitting  arises with gauge symmetry breaking, at tree level,  in the supermultiplets of the Z vector boson and a new X vector boson.  It  spreads to the other particles at one loop. This mass splitting is possible because the exotic invariant changes the algebra of SUSY.  This paper E7 discusses  a number of issues that arise in the calculation of the mass splitting  of the XM.  That mass splitting will require computer programs for solution.

 \Large

 \section{Introduction}
\la{intro}

\subsection{The XM and the ZX sector}

The Exotic Model (``the XM") contains a new mechanism that splits the  mass multiplets in the \SSM\ (``the SSM"), without using  spontaneous or explicit supersymmetry breaking.  The details are set out in E6 \ci{E6}. A special  exotic invariant is added to the SSM, so that the VEVs that break gauge symmetry also split SUSY at tree level. The choice of the \ei\ is  quite unique, because the \ei\ must  satisfy the   constraint equation demanded by the spectral sequence, as set out in E2 \ci{E2}. These two features are linked to the existence of the two Higgs doublets in the SSM.

When the SSM suffers its familiar gauge symmetry breaking from \sg\ to \bsg,  the  neutral charge sector formed from the Z and X vector bosons, and their superpartners,  gets a SUSY violating mass mixing, at tree level.  This happens because the \ei s depend explicitly on pseudofields, and so they change the algebra of supersymmetry.  At tree level, this mass mixing does not affect the masses of the other supermultiplets, which are the quarks, leptons,  $W^+,W^-$, and Higgs and photon.  These all remain in their mass-degenerate supermultiplets, at tree level.

To proceed to the one-loop level we need the propagators that come from the tree level quadratic actions.  For the quarks etc. this is straightforward, but for the ZX multiplet this must be calculated, and that is the subject of this paper E7.

 The paper E6 set down the detailed form of the special \ei\ and the relevant parts of the SSM, including the effect of the shifts of the two Higgs doublet chiral superfield scalars $H^i\ra H^i + m h^i  , K^i\ra K^i + m k^i$.  Naively, all one needs to do is collect the terms and invert them to find the propagators, and then the mass spectrum  should reveal the masses, as the positions of the poles in the propagators as a function of momentum.  Essentially the squared masses $m^2$ could be expected to  appear in the propagators in the form  $\fr{1}{p^2+m^2}$ where $p^2= {\vec p}^2- p_0^2$.  We would expect to   need to diagonalize the mass matrix, as we do in the \SM.  In the \SM\ we need to diagonalize mass matrices for the three flavours of quarks and leptons of given charges.  The quarks (and the leptons) occur in three left SU(2) doublets and six right SU(2) singlets.  They differ only in their coefficients and the CKM $3 \times 3$ matrix algebra will diagonalize their masses after gauge symmetry breaking \ci{quarkmasses}. The diagonalization is the same for the SSM.  

However, this is not so simple for the ZX sector of the XM. There are four Weyl fermions in the ZX multiplet and their   coupled is complicated by the fact that one of them\footnote{The dimensions of the two fermions in the \cdss\ (the ``CDSS")  is evident from the form of the actions in section \ref{tachyon}.}  has dimension $\fr{1}{2}$ while three of them have the usual dimension $\fr{3}{2}$.  So diagonalizing their masses is not simple.   The ZX bosons are also complicated.  This arises because the various particles that get mixed come from different multiplets. Furthermore the X multiplet is not similar to the familiar chiral or vector multiplets.  

In this paper we will discuss some of the difficulties that must be overcome when we want to find the masses and propagators in the ZX sector.  

The first issue is that the  \cdss\ (the ``CDSS"), which is an essential part of the XM,  has tachyonic degrees of freedom.  Fortunately for the XM, however, it appears that the tachyons can be removed, in a way that is slightly familiar--it resembles the removal of tachyons by the Higgs mechanism.  That question is discussed in section \ref{tachyon} below.  There are also many other complications  that relate to this tachyon issue.  A summary of this paper follows in subsection
\ref{summary}.

\subsection{A brief description of the sections of this paper}
\la{summary}

This following  15 points are a summary, in order,  of the 15 sections of this paper:

\bitem
\item
Section \ref{intro} is the present Introduction.
\item
Section \ref{tachyon} discusses the tachyon problem. The \cdss\ is a new and necessary addition to the SUSY representations in the SSM. The crucial issue for the  \cdss\ (the ``CDSS")    is a need for, and a mechanism for, tachyon removal.    
\item
 Section \ref{quad} collects the ZX sector from E6.
As discussed in E6 \ci{E6},  the XM contains a free massive quadratic action which appears at tree level after gauge symmetry breaking from  \sg\ to \bsg.  
 This is the ``ZX sector", which splits SUSY.  The basic question here is: `What is the mass spectrum of the ZX sector?'   

\item
Section \ref{Pseudoquad}  discusses  the Pseudofield action for the ZX sector.

\item
 Section \ref{MEquad} introduces the \ME\ for the ZX sector.  
\item
Section \ref{solforquad} contains the solution for the \ME\ in section \ref{MEquad}.
\item
Section \ref{fermionmatrix} contains the matrix of mixings for the fermions in the ZX sector, based on the results in section \ref{solforquad}.
\item
Section \ref{bosonmatrix} contains the matrix of mixings for the bosons in the ZX sector, based on the results in section \ref{solforquad}.

\item
 Section \ref{diagrams} contains Feymnan  diagrams which contribute to  new counterterms.  These  arise from renormalization of the quadratic action as it appears in E6.  Naively,  the next logical step after sections   \ref{fermionmatrix} and \ref{bosonmatrix}  would be to invert the matrices in   \ref{fermionmatrix} and  \ref{bosonmatrix} to find the propagators, and then the new masses would be the poles in those propagators.
 But the new counterterms cast doubt on the matrices in sections \ref{fermionmatrix} and  \ref{bosonmatrix}. 
  
\item Section \ref{cantrans} discusses how
the new counterterms in section  \ref{diagrams}  should be obtainable using  canonical transformations.  
\item
Section \ref{coding} sets down  a number of coding problems.  These will decide various issues that need detailed calculations.  

\item
Section \ref{specseq} discusses some issues about the spectral sequence.

\item
Section \ref{badintegral} discusses an old and  very general, but incorrect, paper that needs to be corrected.
\item
Section \ref{discuss} discusses the issues that arise from the  tachyon problem of section \ref{tachyon} and the canonical  transformations of section \ref{cantrans}, and the relations between them.  

\item
Section \ref{conclusion} is the Conclusion.   The necessary calculations in this paper are clearly too complicated to be solved, reliably, without using computer programs, and the Conclusion discusses the easiest approach to the mass spectrum of the ZX multiplet.  
\eitem

\section{Nature Abhors a Tachyon.}
\la{tachyon}

The XM  is the SSM augmented with a special \ei, as described in   E6 \ci{E6}.

The most notable, and worrying, feature of the XM is the presence of the \cdss\ multiplet   (the ``CDSS").   There are three parts to its action: 
 \be
\cA_{\rm CDSS}=  
\cA_{1}    
+
\cA_{2}    
+
{\ov \cA}_{2}    
  \ee
where  
 \be
\cA_{1}    
 = r_1\int d^4 x   
\lt \{  {\f}_{\dot \b} \pa^{\a \dot \b} \Box {\ov \f}_{\a} 
+ {X}_{\a \dot \b } \pa^{\a \dot \d} \pa^{\g \dot \b}   {\ov X}_{\g \dot \d} 
+  {\c}_{\dot \b} \pa^{\a \dot \b}  {\ov \c}_{\a}  
\rt \}   \ee
\be 
\cA_{2}    
 =    r_2  \int d^4 x   
\lt \{  - \f^{\dot \b}  (\Box + M^2)     \c_{\dot \b}  
+ \fr{1}{2}   X_{\a \dot \b }      (\Box + M^2)    X^{\a \dot \b}
\rt \}  
\ee
 The  bosonic equations of motion   at lowest order are:

 \be
\fr{\d \cA_{\rm CDSS}}{\d  {X}^{\a \dot \b}}= r_1\pa_{\a \dot \d} \pa_{\g \dot \b}   {\ov X}^{\g \dot \d} 
 +   r_2 (\Box + M^2)    X_{\a \dot \b }=0
 \ee
and
 \be
\fr{\d \cA_{\rm CDSS}}{\d  {\oX}_{\g \dot \d}}= 
  r_1 \pa^{\a \dot \d} \pa^{\g \dot \b}   {X}_{\a \dot \b} 
+
    \ov r_2  (\Box + M^2)    \oX^{\g \dot \d}=0
\ee
and the latter means that
 \be
- \fr{1}{    (\Box + M^2)   } \pa^{\ve \dot \d} \pa^{\g \dot \z} \fr{r_1}{\ov r_2}  {X}_{\ve \dot \z} 
 =
      \oX^{\g \dot \d}
\ee
and so by substitution we have
 \be
\lt \{ r_1^2  \pa_{\a \dot \d} \pa_{\g \dot \b}  \pa^{\ve \dot \d} \pa^{\g \dot \z}   {X}_{\ve \dot \z} 
 -    r_2 \ov r_2 (\Box + M^2)^2  \rt \} X_{\a \dot \b }=0
 \ee
which expands to
 \be
\lt \{\lt ( \fr{r_1^2 }{  r_2 \ov r_2}  -1\rt )\Box^2
 -  2\Box  M^2  -    M^4  \rt \} 
X_{\a \dot \b }=0
 \ee
Having four derivatives like $\Box^2$ in the equation of motion is unacceptable, because it means there are tachyons in the theory\footnote{For example if we took $r_1=1, \rm r_2\ra 0,\rm r_2 M \ra M$ we would get the equation of motion  $\lt (  \Box^2 -    M^4  \rt )X_{\a \dot \b }=0$ which clearly has a solution with imaginary mass. }. 

Now here is the point.  If we choose the parameters so that 
\be
\lt ( \fr{r_1^2 }{  r_2 \ov r_2}  -1\rt )
=0
\ee
this reduces to
 \be
  M^2 \lt (   \Box      +   \fr{ M^2}{2}  \rt ) 
X_{\a \dot \b }=0
 \ee
So by choosing the parameters so that $\lt ( \fr{r_1^2 }{  r_2 \ov r_2}  -1\rt )=0$, we have eliminated the tachyonic mass and the higher derivatives in the lowest order equation of motion. The same choice works for the fermions here. It will be seen later that this choice leads to non-renormalizable terms in the action, but we will accept that for the time being. Our argument is that having non-renormalizable terms is to be expected anyway in the theory, given that supergravity and gravity are non-renormalizable. 

The first question for us in this paper is whether we can still repeat this choice for the full theory with the ZX sector that arises for the XM.  If we can, then that theory will continue to be of some interest, because it will be a theory with supersymmetry splitting, and no tachyons.  

Why would the theory choose parameters so that 
$\lt ( \fr{r_1^2 }{  r_2 \ov r_2}  -1\rt )
=0$?   Recall that the Higgs mechanism is also a mechanism that removes tachyons, in that case by developing VEVs and shifts for scalar fields \ci{quarkmasses}.
 Perhaps it is possible to explain the choice above  in terms of the stability of the vacuum against negative energy excitations.

 \section{The Field Action for the ZX 
Sector}

\la{quad}

\subsection{Some new definitions for the ZX sector}
\la{newfields}

The natural procedure when confronted with the action in E6 is to collect all the quadratic kinetic and mass terms in the ZX sector.  That is easy to do.  We will not display the result of that directly here,  however. We will display a closely related result, which arises if we do a  collection based on the fields which would be natural to the  SSM, if it were not accompanied by the special \ei:
\ben
\item
There is a mixture in the SSM which gives rise to the Z vector boson and its superpartners.  The ZX sector here will be written in terms of 
the  new vector multiplet fields
\be
U_{\a \dot \b}, D, \lam_{\a} 
\ee
whose complex conjugates are
\be
U_{\b \dot \a}= \lt ( { U}_{\a \dot \b}\rt )^* ,D,  {\ov\lam}_{\dot \a}
\ee
and the new chiral multiplet fields
\be
F,A, \y_{\a}
\ee
whose complex conjugates are
\be
\oF,\A, \oy_{\dot \a}
\ee
These new multiplets contain the fields which would appear in the $Z_{\a \dot \b}$  vector boson supermultiplet, in the SSM, after gauge symmetry  breaking\footnote{This unbroken SUSY result in the SSM is a little unusual to see in the literature, since, of course,  terms are usually added to break supersymmetry.  
The $U_{\b \dot \a}$ vector boson is made from the charge neutral U(1) and SU(2) vector bosons. Similarly the auxiliary  $D$ is made from  the charge neutral U(1) and SU(2) auxiliary D fields.  The spinor $\lam_{\a}$ is made from the charge neutral U(1) and SU(2) spinor fields and the spinor $\y$ is one linear combination of the two neutral spinors $\y_H,\y_K$, which also joins the Z multiplet. The other combination of these neutral spinors $\lam_{\a}$ from the vector multiplets gives rise to the photon and the photino which both remain massless at tree level.  The scalar $A$ is made from the charge neutral parts of the H and K scalar doublets. Recall that the Z vector boson mass results from the Higgs mechanism where a scalar gets `eaten'.  The other piece of that complex scalar gets a mass equal to that of the Z from its mixture with the D that arises with the Z. The other complex neutral scalar joins the Higgs multiplet in the SSM with the other linear combination of the two neutral spinors $\y_H,\y_K$.  }, if the SSM were not accompanied by the special \ei\ to form the XM.

 But we select these new names, since in the XM, we can expect that the observed $Z_{\a \dot \b}$  vector boson will be composed of  $U_{\a \dot \b}$ and the complex vector boson $X_{\a \dot \b}$, with corresponding mixtures for the other fields in the Z supermultiplet of the SSM. 

\item
In the XM, all these new fields get mixed with the fields from the CDSS. The fields in the CDSS are the complex $X_{\a \dot \b}$ vector boson and the two complex spinors $\f_{\dot \a},\c_{\dot \a}$. We continue to use the spinors  $\f_{\dot \a},\c_{\dot \a}$, but the vector boson needs to be divided into two components
\be
X_{\a \dot \b} = \fr{1}{\sqrt{2}}\lt ( \cR_{\a \dot \b} +i \cJ_{\a \dot \b}\rt )
\ee
We will return to this later in this paper.

\een

\subsection{Field Action for the Quadratic Action in the Z Sector, in terms of the new fields}

 \la{zxaction}
 
The terms in the ZX sector in E6, with the new notation of subsection \ref{newfields}, can be written in terms of five subactions as follows.  The reason for this follows in sections \ref{Pseudoquad}, \ref{MEquad} and  \ref{solforquad}.

Here are the five independent field actions  $\cA_i $ for the Quadratic Action in the ZX Sector:
 \be
\cA_{1}    
 = r_1\int d^4 x   
\lt \{  {\f}_{\dot \b} \pa^{\a \dot \b} \Box {\ov \f}_{\a} 
+ {X}_{\a \dot \b } \pa^{\a \dot \d} \pa^{\g \dot \b}   {\ov X}_{\g \dot \d} 
+  {\c}_{\dot \b} \pa^{\a \dot \b}  {\ov \c}_{\a}  
\rt \}   \ee
\be 
\cA_{2}    
 =    r_2  \int d^4 x   
\lt \{  - \f^{\dot \b}  (\Box + M^2)     \c_{\dot \b}  
+ \fr{1}{2}   X_{\a \dot \b }      (\Box + M^2)    X^{\a \dot \b}
\rt \}  
\ee
 \be 
\cA_{3}    
 = 
r_{3}  \int d^{4}x \left \{ -\frac{1}{4} 
\lt (\pa_{\m} U_{\n} 
-\pa_{\n} U_{\m} \rt ) \lt ( \pa^{\m} U^{ \n} 
-\pa^{\n} U^{ \m} \rt )
\ebp - \frac{1}{2}  \lambda^{ \a}
 \pa_{\a \dot \b}  {\ov  \lambda}^{ \dot \b} 
 + \frac{1}{2}
D^2  \rt \}
\ee
Then we have the usual Ghost and Gauge Fixing Term that is generated by the BRS variation $\d$ defined below acting on the following expression:
\be 
\cA_{4}    
    = 
 r_{4}  \d   \int d^{4}x \;  \h^{a} \left \{ \frac{1}{2} Z + \fr{1}{2 \a_{\rm g}}\partial^{\a \dot \b}U_{\a \dot \b} 
  + 2m_?  \a_{\rm g} i (A - \A)
\right \}
\ee
The mass $m_?$ is chosen so that the cross term
eliminates the mixing term that arises in the SSM, in the standard way that is used in the Standard Model \ci{quarkmasses}. Here $\a_{\rm g}$ is the gauge parameter for the Z boson.

Finally we write down the mixing parts of the ZX action, in terms of the new variables:
   \be  \cA_{5}    
=
 r_{5} \int d^{4}x  \left \{\cL_{\rm Bose}\rt\}
+ r_{5} \int d^{4}x  \left \{\cL_{\rm Fermi}\rt\}
\la{actionA5}
\ee
where
\be
\cL_{\rm Bose}
=
e_1  F   {\ov F}
 + 
e_2 \pa^{ \a \dot \a  } \A
  \pa_{ \a \dot \a  }   A 
+ e_3 i  \fr{m g}{2} 
\pa^{ \a \dot \a}U_{\a \dot \a }     
   A 
- \ov e_3 i  \fr{m g}{2} 
\pa^{ \a \dot \a}U_{\a \dot \a }     \A
\eb
-  e_4  \lt( \fr{m g}{2} \rt)^2
   U^{\a \dot \a }    U_{\a \dot \a }      
-   e_5  \fr{m g}{2} 
    D  A  
-   \ov e_5  \fr{m g}{2} 
    D    \A 
 + e_6
      X_{\dot \a \a}   \oX^{\dot \a \a}  (2 m^2)  
      \eb  
+ e_7 g_{x}    X^{\dot \a \b}    b_{9}   m  \pa_{\b \dot \a} \A
+ \ov e_7 g_{x}    \oX^{\dot \a \b}    b_{9}   m  \pa_{\b \dot \a} A
\eb
+{e_8} m g_{x}    X^{\dot \a \b}  b_{9}   \fr{m g}{2}  U_{\b \dot \a }    
+  {\ove_8} m g_{x}    \oX^{\dot \a \b}  b_{9}   \fr{m g}{2}  U_{\b \dot \a }    
   \ee
   \be
    +    e_9 g_x^2  
   \fr{1}{2}X^{\dot \a \b}   X_{\dot \a \b}   (2 m^2)   
 + \ov e_9  g_x^2  
   \fr{1}{2}\oX^{\dot \a \b}   \oX_{\dot \a \b}   (2 m^2)   
 \ee
and
  \be  \cL_{\rm Fermi}
=
f_1  \y^{ \a}    \pa_{ \a \dot \a  }  {\ov \y}^{  \dot \a} 
+ f_2 i \fr{m g}{2} 
\oy^{\dot \a}  \ov \lambda_{\dot \a}  
- \ov f_2  i \fr{m  g}{2} 
\y^{\a}  \lambda_{\a}  
\eb+ 
 i    f_3   g_{x}   
b_{4}  \fr{m^2 g}{2}  \f^{\dot \a}
  \ov \lam_{\dot \a}  
- i  \ov f_3    g_{x}   
b_{4}  \fr{m^2 g}{2}  \ov\f^{ \a}
  \lam_{\a} 
 \eb +  f_4 \c^{\dot \a} 
b_{11} m   \oy_{\dot \a}+
\ov f_4   \ov \c^{\a} 
b_{11} m   \y_{\a}\lt.
    +f_5 \f^{\dot \a }   \c_{\dot \a}     (2 m^2)  
 +\ov f_5   \ov\f^{\a }   \ov\c_{\a}     (2 m^2)  
  \ebp
  +
f_6 \f^{\dot \a}  \pa_{\a \dot \a} \ov \f^{\a}
  (2 m^2)  
\rt \}
    \ee
    
     Each of these five actions is chosen to be invariant under the BRS transformations in section \ref{MEquad}, as will be discussed in section \ref{solforquad}. 
     The proof of that invariance does require some constraints on the parameters in  $\cA_5$. We have not tried to write down the relation between these coefficients and the ones used in E6.  
There are several problems to solve before we attempt to do that.

 \section{The Pseudofield Action for the ZX  Sector}
\la{Pseudoquad} 
 
The XM is constrained by a \ME, as set out in E6 \ci{E6}.  For the quadratic terms in the ZX sector, this gives rise to a \ME,  set out in terms of the fields of the ZX sector. We will write that down in section \ref{MEquad}. Here are the relevant parts of the Pseudofield action, written in terms of the new fields in subsection \ref{newfields}:
\be
 \cA_{\rm Real \; \Pf s} = \int d^4 x \;
\ee
\be
\left \{ 
\S^{ \a \dot \b } 
\lt [ \pa_{\a \dot \b} \w
  + 
  C_{\a} 
\ov{\lambda}_{ \dot{\b} }   +
  \lambda^{}_{\a } 
{\ov C}_{\dot{\b} }   +  \x^{\g \dot \d}\pa_{\g \dot \d}U_{\a \dot \b} \rt ]
\ebp
+
\D^{}
\lt [\fr{-i}{2}  C^{\a} 
  \pa_{\a \dot \b}  {\ov  \lambda}^{ \dot \b} 
  +\fr{i}{2}\ov{C}^{\dot \b}
  \pa_{\a \dot \b}  { \lambda}^{ \a} 
+
 \x^{\g \dot \d}\pa_{\g \dot \d} D^{}
\rt ]
\ebp
 +
\z
\lt ( C^{\a}  
\ov{C}^{\dot \b}
\pa_{\a \dot \b} \h +  \x^{\g \dot \d}\pa_{\g \dot \d}Z 
\rt )
+
H^{} 
\lt ( Z+  \x^{\g \dot \d}\pa_{\g \dot \d}  \h 
\rt )
\ebp
+
W^{}
\lt [
  C^{\a} \ov{C}^{\dot \b}  U_{\a\dot \b} +
 \x^{\g \dot \d}\pa_{\g \dot \d} \omega\rt ]
\right \}
-  P_{\a \dot \b}
  C^{\a} \ov{C}^{\dot \b}
\ee
and
 \be
\cA_{\rm Complex \; \Pf s}=
\eb
 \int d^4 x \lt \{\G \lt [ \y_{\b} C^{\b} 
-   i \fr{m_{1} g}{2} \w  + b_{1}   g_{x}   m_{2}   \f^{\dot \a}\oC_{ \dot \a}   
+ \x^{\g \dot \d} \pa_{\g \dot \d}    A\rt ]\ebp
+Y^{\a} \lt [
\lt ( \pa_{\a \dot \b}  A +  i \fr{m_{3} g}{2}   U_{\a \dot \b} 
  \rt ) {\ov C}^{\dot \b}   
+ F C_{ \a}+  b_{6}  g_x  m_{4}    X_{\dot \a \a}  \oC^{ \dot \a}  
 + \x^{\g \dot \d} \pa_{\g \dot \d}    \y_{\a  }\rt ]
 \ebp
  + \Lam \lt [
  \pa_{\a \dot \b}  \y^{\a}  {\ov C}^{\dot \b} +  i \fr{m_{5} g}{2}   {\ov \lambda}_{ \dot \b}   {\ov C}^{\dot \b} +   b_{10}  g_x  m_{6} \c^{\dot \a}  \oC_{\dot \a} 
 +\x^{\g \dot \d} \pa_{\g \dot \d}   F \rt ]
\ebp
+ L^{\a}\lt[  \frac{1}{2}\lt (
 \pa_{\a \dot \b} U_{\b}^{\;\; \dot \b}
+ \pa_{\b \dot \b} U_{\a}^{\;\; \dot \b}
 \rt )
   C^{\b}   - i D^{}
C_{\a}  +  \x^{\g \dot \d}\pa_{\g \dot \d}
\lambda^{}_{\a} 
 \rt ]
 \ebp
   +G^{\dot \a}\lt [ C^{\d} X_{\dot \a \d} +\x^{\g \dot \d} \pa_{\g \dot \d} \f_{\dot\a} \rt ] 
   \ebp
    +\X^{\a \dot \b} \lt [ \pa_{ \a \dot \d }  \f_{\dot\b} {\ov C}^{\dot \d}  
+ 
C_{\a}   
\c_{\dot\b} +\x^{\g \dot \d} \pa_{\g \dot \d} X_{\a \dot \b}\rt ] 
\ebp
+  K^{\dot \a} \lt [  \pa^{\d\dot \d}  X_{\dot \a \d }   {\ov C}_{\dot \d}+
\x^{\g \dot \d} \pa_{\g \dot \d}  \c_{\dot \a}  \rt ] 
\rt \} 
\ee
with \CC:
 \be
\ov \cA_{\rm Complex \; \Pf s}=
=
\eb
 \int d^4 x \lt \{\ov\G \lt [ \oy_{\dot\b} \oC^{\dot\b}
+  i \fr{m_1 g}{2} \w  + b_{1}   g_{x}   m_2    \ov \f^{\a} C_{\a}   
+ \x^{\g \dot \d} \pa_{\g \dot \d}    \A\rt ]\ebp
+\oY^{\dot\a} \lt [
\lt ( \pa_{\a \dot \a}  \A  -  i \fr{m_3 g}{2}   U_{\a \dot \b} 
  \rt ) { C}^{\a}   
+ \oF \oC_{\dot \a} 
+  b_{6}  g_x  m_4    \oX_{\dot \a \a}  C^{ \a}  
 + \x^{\g \dot \d} \pa_{\g \dot \d}    \oy_{\dot \a  }\rt ]
  \ebp
  + \ov \Lam \lt [
  \pa_{\b \dot \a}  \oy^{\dot\a}  { C}^{\b} -  i \fr{m_5 g}{2}  
   { \lambda}_{ \b}   {C}^{\b} +   b_{10}  g_x  m_6 \ov \c^{ \a}  C_{\a} 
 +\x^{\g \dot \d} \pa_{\g \dot \d}   \oF \rt ]
\ebp
+ \oL^{\dot\a}\lt[  \frac{1}{2}\lt (
 \pa_{\b \dot \a} U_{\;\;\;\dot \b}^{\b}
+  \pa_{\b \dot \b} U_{\;\;\;\dot \a}^{\b}
 \rt )
   \oC^{\dot \b}   + i D^{}
\oC_{\dot\a}  +  \x^{\g \dot \d}\pa_{\g \dot \d}
\ov\lambda_{\dot \a} 
 \rt ]
 \ebp
   +\ov G^{ \a}\lt [ \oC^{\dot\d} \oX_{\a \dot \d} +\x^{\g \dot \d} \pa_{\g \dot \d} \ov\f_{\a} \rt ] 
   \ebp
    +\ov\X^{\a \dot \b} \lt [ \pa_{ \d \dot \b } \ov \f_{\a} {C}^{\d}  
+ 
{\ov \c}_{\a}   
\oC_{\dot\b} +\x^{\g \dot \d} \pa_{\g \dot \d} \oX_{\a \dot \b}\rt ] 
\ebp
+ \oK^{ \a} \lt [  \pa^{\d\dot \d}  \oX_{ \a \dot\d }   {C}_{\d}+
\x^{\g \dot \d} \pa_{\g \dot \d}  \ov \c_{  \a}  \rt ] 
\rt \} 
\ee
These transformations will be chosen to be nilpotent.   Calculation reveals that  nilpotence requires the constraints:
\be
 \fbox{\colorbox{yellow}{ $m_{1} = m_{3}= -m_{5};\;m_{2} =- m_{4} = m_{6}
$}} 
\ee

 \section{The BRS \ME\ for the quadratic ZX sector}

\la{MEquad}

The invariance and the nilpotence described above are both summarized by the ZX Master Equation, which has the following form:
\be
\cM_{\rm Total}= \cM_{\rm Real}+\cM_{\rm Complex}+\ov\cM_{\rm Complex}=0
\ee
where
\be
\cM_{\rm Real}= \int d^4 x \; \lt \{
\fr{\d \cA}{\d \S^{\a\dot\b}} \fr{\d \cA}{\d U_{\a \dot \b}} 
+ \fr{\d \cA}{\d  \D } \fr{\d \cA}{\d D} 
+ \fr{\d \cA}{\d  Z } \fr{\d \cA}{\d \z} 
+ \fr{\d \cA}{\d  H } \fr{\d \cA}{\d \h}+ \fr{\d \cA}{\d   W} \fr{\d \cA}{\d \w} \rt \}
\eb
 + \fr{\pa \cA}{\pa P^{\a \dot \b} }
 \fr{\pa \cA}{\pa \x_{\a \dot \b} }
\ee
and
\be
\cM_{\rm Complex}= \int d^4 x \; \lt \{
\fr{\d \cA}{\d \G} \fr{\d \cA}{\d A} 
+ \fr{\d \cA}{\d  Y^{\a}  } \fr{\d \cA}{\d  \y_{\a}} 
+ \fr{\d \cA}{\d L^{\a}} \fr{\d \cA}{\d  \lam_{\a} } 
\ebp
+ \fr{\d \cA}{\d G^{\dot\a}} \fr{\d \cA}{\d  \f_{\dot\a} } 
+ \fr{\d \cA}{\d K^{\dot\a}} \fr{\d \cA}{\d  \c_{\dot \a} } 
+\fr{\d \cA}{\d \X^{\a\dot\b}} \fr{\d \cA}{\d X_{\a \dot \b}} 
\rt \}
\ee
and

\be
\ov\cM_{\rm Complex}= \int d^4 x \; \lt \{
\fr{\d \cA}{\d \ov \G} \fr{\d \cA}{\d \A} 
+ \fr{\d \cA}{\d  \oY^{\dot\a}  } \fr{\d \cA}{\d  \oy_{\dot\a}} 
+ \fr{\d \cA}{\d \oL^{\dot \a}} \fr{\d \cA}{\d  \ov\lam_{\dot\a} } 
\ebp
\ebp
+ \fr{\d \cA}{\d \ov G^{\a}} \fr{\d \cA}{\d  \ov \f_{\a} } 
+ \fr{\d \cA}{\d \ov K^{\a}} \fr{\d \cA}{\d  \ov \c_{\a} } 
 +\fr{\d \cA}{\d \ov \X^{\a\dot\b}} \fr{\d \cA}{\d \oX_{\a \dot \b}} 
\rt \}
\ee
The action $\cA$ in these expressions is given by
\[
\cA= 
\cA_{\rm Complex \; \Pf s}+
\cA_{\rm Real \; \Pf s}+
\ov \cA_{\rm Real \; \Pf s}
\]\be
- P^{\a \dot \b} C_{\a}\oC_{\dot \b}
+ \cA_1+ \cA_2+ \ov \cA_2+ \cA_3+ \cA_4+ \cA_5
\la{invariantaction}\ee

 \section{The ZX 
Sector as a solution for the its \ME}
\la{solforquad}
The \ME\ in section \ref{MEquad} is satisfied for the expressions in section \ref{quad} and \ref{Pseudoquad}, except that invariance of the expression 
 $\cA_{5}$ in (\ref{actionA5}) requires that all the following equations must hold:
\be
 \left(
\begin{array}{ccc}
1.&e_1+f_1 & e_1+f_1 \\
2.& f_2 m-e_1 m_1 & e_1 m_1-m {\ov f}_2 \\
3.& e_1 m_2 g_x+f_4 m & e_1 m_2 g_x+m {\ov f}_4 \\
4.& 2 e_2+f_1 & 2 e_2+f_1 \\
5.& e_2 m_1-e_3 m & e_2 m_1-m {\ov e}_3 \\
6.& 2 e_2 m_2+m {\ov e}_7 & e_7 m+2 e_2 m_2 \\
7.& e_3+\frac{e_5}{2} & -{\ov e}_3-\frac{{\ov e}_5}{2} \\
8.& e_3-\frac{e_5}{2}-{\ov f}_2 & f_2-{\ov e}_3+\frac{{\ov e}_5}{2} \\
9.& 2 e_4 m+e_3 m_1+m_1 {\ov e}_3 & 2 e_4 m+e_3 m_1+m_1 {\ov e}_3 \\
10.& -2 e_4 m-f_2 m_1 & -2 e_4 m-m_1 {\ov f}_2 \\
11.& -e_5-{\ov f}_2 & f_2+{\ov e}_5 \\
12.& e_5-{\ov e}_5 & e_5-{\ov e}_5 \\
13.& f_3 m-e_5 m_2 & m {\ov f}_3-m_2 {\ov e}_5 \\
14.& f_5-e_9 g_x^2 & {\ov e}_9 g_x^2-{\ov f}_5 \\
15.& 2 e_6 m+f_4 m_2 g_x & 2 e_6 m+m_2 {\ov f}_4 g_x \\
14\;{\rm again}.& e_9 g_x^2-f_5 & {\ov e}_9 g_x^2-{\ov f}_5 \\
16.& f_4-e_7 g_x & {\ov f}_4-{\ov e}_7 g_x \\
17.& e_8 m+i e_7 m_1 & m {\ov e}_8-i m_1 {\ov e}_7 \\
18.& -e_8 m_1 g_x-i f_4 m_1 & -m {\ov e}_8 g_x+i m_1 {\ov f}_4 \\
19.& -e_8-i f_3 & -{\ov e}_8+i {\ov f}_3 \\
20.& -e_8 m+i m_2 {\ov f}_2 & -m {\ov e}_8-i f_2 m_2 \\
\end{array}
\right)=0
\ee
A lengthy and careful calculation shows that the above 20 complex equations have the solution:
 \be
  \left\{m_1\to m,e_2\to \frac{e_1}{2},e_3\to \frac{e_1}{2},{\ov e}_3\to \frac{e_1}{2},
  \ebp
  e_4\to -\frac{e_1}{2},e_5\to -e_1,{\ov e}_5\to -e_1,e_6\to \frac{e_1 m_2^2 g_x^2}{2 m^2},
  \ebp
  e_7\to -\frac{e_1
   m_2}{m},{\ov e}_7\to -\frac{e_1 m_2}{m},e_8\to \frac{i e_1 m_2}{m},{\ov e}_8\to -\frac{i e_1 m_2}{m},f_1\to -e_1,
   \ebp
   f_2\to e_1,{\ov f}_2\to e_1,f_3\to -\frac{e_1 m_2}{m},{\ov f}_3\to
   -\frac{e_1 m_2}{m},f_4\to -\frac{e_1 m_2 g_x}{m},
   \ebp
   {\ov f}_4\to -\frac{e_1 m_2 g_x}{m},f_5\to e_9 g_x^2,{\ov f}_5\to {\ov e}_9 g_x^2\right\}
 \ee
 The invariance also requires that 
 \be
b_{1}=
b_{2}=
b_{3}=
b_{4}=b_{7}=b_{10}=1
;b_{5}=
b_{6}=
b_{8}=
b_{9}=
b_{11}= -1
\la{valuesofbs} 
\ee
which are explained in \ci{E4}.
There are also 16 more equations, for a total of 56 equations.  
The additional  16 equations require a solution for $f_6$, as follows
\be
   f_6  
=-
  e_6  = -\frac{e_1 m_2^2 g_x^2}{2 m^2}
\la{valoff6}
\ee 

This calculation should be repeated using a computer program in order to verify it. 

Here is the point.  
The expressions for the ZX sector in E6 are written in terms of the fields that are natural to the action in that paper.  However because of mixing, as described in subsection \ref{newfields}, those are not the easiest fields to use for the ZX sector. 
The fields we use are fields which are fields which would be eigenstates of mass in the SSM, but they do get mixed in the XM.  They do however form a represenation of the Master 
Equation in section \ref{MEquad}, and that helps us to check that we have them correctly written.

 \section{The Fermionic Field Matrix in the ZX 
Sector}

\la{fermionmatrix}

We can write the free massive action for the fermionic part of the ZX sector, starting from the expressions in section \ref{solforquad}, in the following way:

\be  \cA_{\rm ZX \;Quad \;Fermi}  
\eb
= \int d^4 x 
  \left(
\begin{array}{cccc}
 \f_{\dot \a}&\oy_{\dot \a}&\ov \c_{ \a} &  \lam_{\a} \\
\end{array}
\rt )
\lt (\begin{array}{cc}
M^{\dot \a \b }
&M^{ \dot \a \dot\b}\\
M^{\a \b }
&M^{\a \dot\b  }\\
\end{array}
\rt ) 
\left(
\begin{array}{c}
\ov\f_{ \b}\\ \y_{ \b},\\ \c_{\dot \b}\\  \ov \lam_{\dot \b}\\
\end{array}
\rt )
\ee
This form contains information about the way that the indices of these fermions interact.    
The matrix can be written in more detail as: 
 \be
\lt (\begin{array}{cc}
M^{\dot \a \b }
&M^{ \dot \a \dot\b}\\
M^{\a \b }
&M^{\a \dot\b  }\\
\end{array}
\rt )   =
\eb
  \left(
\begin{array}{cccc}
 \fbox{\colorbox{yellow}{$a_{1}  (\Box +n_{1} m ^2) \pa^{\b \dot \a}$}} 
 &   a_{10}  m  \pa^{\b \dot \a}
& 
  \fbox{\colorbox{yellow}{$a_{5}   (  \Box  + n_{2} m ^2  )   \ve^{\dot \b \dot \a}$}}      
 & 
 i  a_{6}
    ( \Box +n_{3}m^2  )    \ve^{\dot \b \dot \a}
  \\
  a_{10}  m  \pa^{\b \dot \a}&   \fbox{\colorbox{yellow}{$a_{3}  \pa^{\b \dot \a} $}}   
&  a_{7}  m   \ve^{\dot \b \dot \a}  
&\fbox{\colorbox{yellow}{$     i  a_{8}  m    \ve^{\dot \b \dot \a} $}}   
  \\
\fbox{\colorbox{yellow}{$  a_{5}  (  \Box  + n_{2} m ^2  )   \ve^{\b\a} $}}   
  &  a_{7}  m   \ve^{ \b   \a}  &  
  \fbox{\colorbox{yellow}{$ 
a_{2} \pa^{\a \dot \b} $}}   
& i  a_{9}  \pa^{\a \dot \b}
 \\
- i  a_{6}
    ( \Box +n_{3}m^2  )  \ve^{ \b   \a}  
&\fbox{\colorbox{yellow}{$   - i  a_{8}  m  \ve^{ \b   \a}  $}}   
&- i  a_{9}  \pa^{\a \dot \b}& 
\fbox{\colorbox{yellow}{$  a_{4}  \pa^{\a \dot \b}  $}}   
  \\
\end{array}
\rt )
\la{fermionmatrixeq}\ee
To generate the propagator we must find a matrix $\lt (
\begin{array}{cc}
{\rm In}_{ \b \dot \g}
&{\rm In}_{ \b \g}\\
{\rm In}_{\dot\b   \dot \g}
&{\rm In}_{\dot\b  \g}\\
\end{array}
\rt )  $ which satisfies 
 \be
\lt (\begin{array}{cc}
M^{\dot \a \b }
&M^{ \dot \a \dot\b}\\
M^{\a \b }
&M^{\a \dot\b  }\\
\end{array}
\rt ) 
\lt (
\begin{array}{cc}
{\rm In}_{ \b \dot \g}
&{\rm In}_{ \b \g}\\
{\rm In}_{\dot\b   \dot \g}
&{\rm In}_{\dot\b  \g}\\
\end{array}
\rt )  
=\lt( \begin{array}{cc}
\d^{ \dot \a}_{  \dot \g}&0\\
0 &   \d^{   \a}_{    \g} \\
\end{array}
\rt )  \equiv
  \left(
\begin{array}{cccc}
\d^{\dot \a}_{\dot \g}
&0&0&0
\\
0&\d^{\dot \a}_{\dot \g}
&0&0
\\
0&0&\d^{ \a}_{ \g}
&0
\\
0&0&0&\d^{ \a}_{ \g}
\\
\end{array}
\rt )
\ee
In a more detailed form this is:
\be
\lt (
\begin{array}{cc}
{\rm In}_{ \b \dot \g}
&{\rm In}_{ \b \g}\\
{\rm In}_{\dot\b   \dot \g}
&{\rm In}_{\dot\b  \g}\\
\end{array}
\rt )  
=  \left(
\begin{array}{cccc}
 t_{1}  \pa_{\b \dot \g}   &  t_{2}    \pa_{\b \dot \g}& 
    t_{3}     \ve_{\b  \g}
 & 
 i  t_{4} 
    \ve_{\b  \g}
  \\
 t_{2}  \pa_{\b \dot \g}   &  t_{5}    \pa_{\b \dot \g}& 
    t_{6}     \ve_{\b  \g}
 & 
 i  t_{7} 
    \ve_{\b  \g}
  \\
  t_{3}    \ve_{\dot \b\dot \g}  &  t_{6}    \ve_{\dot \b\dot \g}  &  t_{8} \pa_{\g \dot \b}& i  t_{9}  \pa_{\g \dot \b}
  \\
-i  t_{4}    \ve_{\dot \b\dot \g}  & -i t_{7}    \ve_{\dot \b\dot \g}  &-i  t_{9} \pa_{\g \dot \b}&  t_{10}  \pa_{\g \dot \b}
 \\
\end{array}
\rt )
\ee

Note that the above forms contain new coefficients.  Also we have not tried to find the relation between these coefficients and the ones used in section \ref{solforquad}.  Nor have we tried to write down the relations between those coefficients and the ones used in E6. In addition we have added some new terms in the above, which could arise as explained in section \ref{diagrams} and \ref{cantrans} below.  These forms are set up for a solution using a computer program, which is currently being developed.  They are clearly too complicated to do otherwise.  Also, as explained below in section \ref{discuss}, we need to subject the equations here to  tachyon equations like those discussed in section \ref{tachyon}. One crucial question is whether it is possible to eliminate all the tachyons.

 \section{The Bosonic Field Matrix in the ZX 
Sector}

\la{bosonmatrix}

Put 
\be
X^{\a \dot \b} =\fr{1}{\sqrt{2}}\lt ( \cR^{\a \dot \b} + i \cJ^{\a \dot \b} 
\rt )
\ee
\be
\oX^{\a \dot \b} =\fr{1}{\sqrt{2}}\lt ( \cR^{\a \dot \b} - i \cJ^{\a \dot \b} 
\rt )
\ee
\be
\cR^{\a \dot \b} =\fr{1}{\sqrt{2}}\lt ( X^{\a \dot \b} +  \oX^{\a \dot \b} 
\rt )
\ee
\be
\cJ^{\a \dot \b} =\fr{i}{\sqrt{2}}\lt (  - X^{\a \dot \b} +\oX^{\a \dot \b}
\rt )
\ee

These are Hermitian in the sense that
\be
(\cR^{\a \dot \b})^*= \cR^{\b \dot \a}
\ee
\be
(\cJ^{\a \dot \b})^*= \cJ^{\b \dot \a}
\ee
but the original vector is not Hermitian:
\be
(X^{\a \dot \b})^*= \oX^{\b \dot \a}
\ee
We also define
\be
A= \fr{1}{\sqrt{2}}\lt ( S+ i G\rt )
\ee
\be
\A= \fr{1}{\sqrt{2}}\lt ( S - i G\rt )
\ee
and we
define \be
m_g=  \fr{m g}{2} 
\ee
and
\be
m_x=   m g_x
\ee

Now we can write the free massive quadratic bosonic terms of the ZX sector in the form:
\be  \cA_{\rm ZX \;Quad \;Bose}  
= \int d^4 x  \cL
\ee
where
\be
 \cL= \left(
\begin{array}{cccccc}
 \cR_{ \a \dot \b}& S&\cJ_{ \a \dot \b}  & U_{ \a \dot \b} & G & \h \\
\end{array}
\rt) \cM_{\rm ZX\; Bosons}   
 \left(
\begin{array}{c}
 \cR^{ \g \dot \d}\\ 
  S \\ 
   \cJ^{ \g \dot \d}\\ 
 U^{ \g \dot \d}\\ 
 G \\ 
 \w \\ 
\end{array}
\rt)\ee
Here the matrix $ \cM_{\rm ZX\; Bosons}$ has the form:
\small
\be
  \left(
\begin{array}{cccccc}
\begin{array}{c} \fbox{\colorbox{yellow}{$ a_1\pa^{\a \dot \b}  \pa_{\g \dot \d} + \d^{\a}_{\g} \d_{\dot \d}^{\dot \b}$}} 
 \\
 \fbox{\colorbox{yellow}{$ a_2 (\Box+m_0^2 +m_x^2 ) $}} \\
\end{array} & 
\fbox{\colorbox{yellow}{$m_x \pa^{\a \dot \b}  $}}      
& 
0 &  
  blip
& 
  0& 
  0
 \\
\fbox{\colorbox{yellow}{$ m_x \pa_{\g \dot d}$}}  
&
\fbox{\colorbox{yellow}{$ \Box + m_g^2$}}& 0   & 0    &0 
&0   
  \\
  0&  0&
 \begin{array}{c} \fbox{\colorbox{yellow}{$\pa^{\a \dot \b}  \pa_{\g \dot \d}  $}}\\
\fbox{\colorbox{yellow}{$+ \d^{\a}_{\g} \d_{\dot \d}^{\dot \b} (d_2-d_1)\Box$}} 
 \\
 \fbox{\colorbox{yellow}{$ (+ m_0^2 -m_x^2) $}} \\
\end{array} 
&  \fbox{\colorbox{yellow}{$m_x^2 \d^{\a}_{\g} \d_{\dot \d}^{\dot \b}  $}}      

 &\fbox{\colorbox{yellow}{$ m_x   \pa^{\a \dot \b} 
 $}}   & 
  0

  \\
  0&    0&  
\fbox{\colorbox{yellow}{ $m_x^2 \d^{\a}_{\g} \d_{\dot \d}^{\dot \b}   $}}   
  &  
  \begin{array}{c} \fbox{\colorbox{yellow}{$(-1+\fr{1}{2\a} )\pa^{\a \dot \b}  \pa_{\g \dot \d} $}} \\
 \fbox{\colorbox{yellow}{$ -\fr{1}{2}  (\Box+m_g^2)
\d^{\a}_{\g} \d_{\dot \d}^{\dot \b}$}} \\
\end{array} 
&0
&0
  \\
0  &0  & \fbox{\colorbox{yellow}{$ m_x \pa^{\g\dot\d}  $}}  &0 
& 
\fbox{\colorbox{yellow}{$  \Box +2 \a m_g^2$}}   
&0    \\
0  
&0   
&0& 0& 0& 
\fbox{\colorbox{yellow}{$ \Box +2 \a m_g^2$}}   
  \\
\end{array}
\rt )
\ee
\Large

\Large
We have included the ghost and antighost terms here just for completeness. The cross term in the Ghost and Gauge Fixing action is used to eliminate one of the terms in this m\
atrix. We can choose a block matrix form for each of these three blocks, and require that the matrix times its inverse matrix results in a unit matrix.
The inverse   necessarily has a form with three corresponding blocks: 
\be    \cM_{\rm ZX \;Bosons}^{-1}=
\ee
\be
  \left(
\begin{array}{cccccc}
\begin{array}{c} \fbox{\colorbox{yellow}{$ t_{1} \pa^{\g \dot \d}  \pa_{\e \dot \z} $}} 
 \\
 \fbox{\colorbox{yellow}{$ +z_{1}   \d^{\g}_{\e} 
 \d^{\dot \d}_{\dot \z} $}} \\
\end{array}  
&\fbox{\colorbox{yellow}{$t_{2} \pa^{\g \dot \d}$}} &0&0 &0&0  
 \\ \\
\fbox{\colorbox{yellow}{$- t_{2} \pa_{\e \dot \z}$}}   &   \begin{array}{c}  
 \fbox{\colorbox{yellow}{$z_{2}$}} \\
\end{array}  
&0 &0 &0 &0  
\\ \\ 

0 & 0 & 
\begin{array}{c} \fbox{\colorbox{yellow}{$ t_{3} \pa^{\g \dot \d}  \pa_{\e \dot \z} $}} 
 \\
 \fbox{\colorbox{yellow}{$ +z_{3}   \d^{\g}_{\e} 
 \d^{\dot \d}_{\dot \z} $}} \\
\end{array}  
&
\begin{array}{c} \fbox{\colorbox{yellow}{$ t_{4} \pa^{\g \dot \d}  \pa_{\e \dot \z} $}} 
 \\
 \fbox{\colorbox{yellow}{$ +z_{4}   \d^{\g}_{\e} 
 \d^{\dot \d}_{\dot \z} $}} \\
\end{array}  &\fbox{\colorbox{yellow}{$ t_{5} \pa^{\g \dot \d}$}}&0  
 \\ \\ 
0 &0 &\begin{array}{c} \fbox{\colorbox{yellow}{$ t_{4} \pa^{\g \dot \d}  \pa_{\e \dot \z} $}} 
 \\
 \fbox{\colorbox{yellow}{$ +z_{4}   \d^{\g}_{\e} 
 \d^{\dot \d}_{\dot \z} $}} \\ 
\end{array}   &
\begin{array}{c} \fbox{\colorbox{yellow}{$ t_{6} \pa^{\g \dot \d}  \pa_{\e \dot \z} $}} 
 \\
 \fbox{\colorbox{yellow}{$ +z_{6}   \d^{\g}_{\e} 
 \d^{\dot \d}_{\dot \z} $}} \\
\end{array}  
&\fbox{\colorbox{yellow}{$ t_{7} \pa^{\g \dot \d}$}} &0  
\\  \\ 
0 &0&\fbox{\colorbox{yellow}{$ -t_{5} \pa_{\e \dot \z}$}}&\fbox{\colorbox{yellow}{$- t_{7}\pa_{\e \dot \z}$}} &   \begin{array}{c}  
	 \fbox{\colorbox{yellow}{$   z_{7} $}} \\
\end{array}  
&0    \\ \\ 
0 &0&0&0 &   &  \begin{array}{c}  
 \fbox{\colorbox{yellow}{$   z_{8} $}} \\
\end{array}  
    \\
\end{array}
\rt )
\ee
\Large
and we define the latter by
\be    M_{\rm ZX\;Bosons}  M_{\rm ZX\; Bosons}^{-1}= {\rm Unity} 
\ee
where
\be
 {\rm Unity}=
  \left(
\begin{array}{cccccc}
\begin{array}{c}  
 \fbox{\colorbox{yellow}{$  \d^{\a}_{\e} 
 \d^{\dot \b}_{\dot \z} $}} \\
\end{array}  
&0 &0&0 &0&0   \\
0 &\begin{array}{c}  
\begin{array}{c}  
 \fbox{\colorbox{yellow}{$  1 $}}\\
\end{array}  
 \\
\end{array}  
&0&0 &0&0   \\
0 &0 &\begin{array}{c}  
 \fbox{\colorbox{yellow}{$  \d^{\a}_{\e} 
 \d^{\dot \b}_{\dot \z} $}} \\
\end{array}  
&0 &0&0  \\
0 &0&0&   \begin{array}{c}  
 \fbox{\colorbox{yellow}{$  \d^{\a}_{\e} 
 \d^{\dot \b}_{\dot \z}$}} \\
\end{array}  
&0 &0\\
0 &0&0&0 &   \begin{array}{c}  
	 \fbox{\colorbox{yellow}{$  1 $}} \\
\end{array}  
&0    \\
0 &0&0&0 &0   &  \begin{array}{c}  
 \fbox{\colorbox{yellow}{$  1$}} \\
\end{array}  
    \\
\end{array}
\rt )
\ee

In this section, we have not tried to indicate the changes that arise when the theory includes terms that can arise from the considerations in section \ref{diagrams} and \ref{cantrans}.  The above is meant to reflect what we get from E6 and section \ref{zxaction}. Again we do not attempt here to relate the coefficients in E6 to those in section \ref{zxaction} or section \ref{solforquad} to  the coefficients here.  Those calculations are left for the coding work. Again here, it is essential that we eliminate the tachyons and the higher derivatives.  Can that be done sensibly?

\Large

 \section{Diagrams for One Loop Counterterms that add to the ZX Sector}
    
\la{diagrams}

It is natural to look at the action in E6 to see whether there are new terms generated at one loop by renormalization.  Here we are generating new counterterms in a complicated action governed by a \ME. It is easy to see that new terms do probably get generated.  

Our hope in this paper is that this exercise does not need to be done here.  So we will not try to find all the term that could arise.  We will simply suggest what could happen.

This is a subject with a venerable history.  The original papers on the subject were written in the context of the renormalization of composite operators \ci{ Kluberg-Stern:1974iel,  Kluberg-Stern:1975ebk, Kluberg-Stern:1974nmx,Deans:1978wn,Dixon:1974ss,Dixon:1975si}.

  A confusing feature is that some of the propagators that we need ro use to find these counterterms seem, in fact, to be the very propagators that we are looking for in the ZX sector.  
Usually the composite operators did not help to  generate the propagators that determine them, but here they might,  and the tachyon-free requirement might complicate matters. 

\subsection{The new counterterm  $  {\cal G}[\f,\y ]\Ra \cA_{\rm New\; Counterterm} = \int d^4 x \lt \{ 
m  \;     \f \cdot\pa  \cdot \y
\rt \} $}

We are picturing the use of the new variables in section \ref{newfields} here.  That is appropriate  after the VEVs have appeared.  But at this time we are not looking carefully at the resulting action or the following diagrams. These diagrams are just for illustration of what could happen in a more detailed study.  

%\normalsize
%template-improved March 1, 2000:
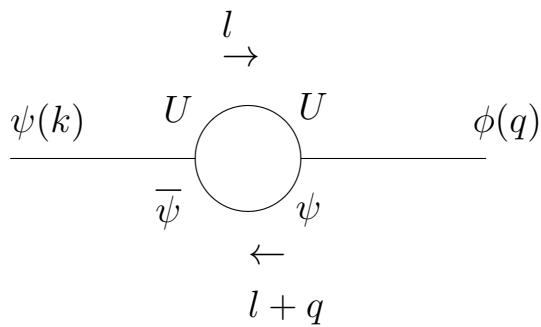
\begin{picture}(250,150)
%\put(100,02){Feynman Diagram for ${\cal G}_1[\f, L, \oy_{K}] ; k+q=0$ }
%
%
%
%\put(0,150){ $ integral $  }
%
% four point at left:
%\put(180,75){\line(-1,-1){50}}
%\put(180,75){\line(-1,1){50}}
% top oblique at right:
%\put(210,92){\line(1,2){15}}
% bottom oblique at right:
%\put(210,58){\line(1,-2){20}}
% horizontal at right:
\put(220,75){\line(1,0){70}}
% four point at right:
%\put(220,75){\line(1,1){50}}
%\put(220,75){\line(1,-1){50}}
% horizontal at left:
\put(180,75){\line(-1,0){70}}
\put(200,75){\circle{40}}
%
% outer fields
%a1
%\put(95,20){${\ov a}_1(p)$}
%a2
%\put(95,120){${\ov a}_2(p)$}
%a3
%\put(230,120){$\f^{\dot \a}(0)$}
%a4
%\put(280,120){${\ov a}_4(p)$}
%a5
\put(285,85){$\f(q)$ }
%a6
%\put(280,20){${\ov a}_6(q)$}
%a7
%\put(235,20){${\ov \lam}_{ \dot \a}(q)$}
% inner Fields:
%w
\put(110,85){{$\y(k)$}}
%b1
\put(168,89){$U $}
%b2
%\put(190,99){${\ov\y}$}
%b4
\put(219,90){$U$}
%b5
\put(218,52){$\y$}
%b7
%TT\put(195,42){${\ov L}$}
%b8
\put(165,51){$\oy$ }
% masses
%m1
%\put(186,88){$\ast$}
%\put(184,81){${m }$}
%m2
%\put(207,89){$\ast$}
%\put(199,84){${m}$}
%m3
%\put(217,73){$\ast$}
%\put(201,72){${m_3}$}
%m4
%\put(207,55){$\ast$}
%\put(200,64){${m}$}
%m5
%\put(186,56){$\ast$}
%\put(185,64){$m_5$}
%m6
%\put(177,73){$\ast$}
%\put(185,73){${m}$}
%
% momenta
%l-p
\put(190,110){$\rightarrow$}
\put(190,121){$l$}
%l
%\put(250,75){$\downarrow$}
%\put(265,75){$l$}
%l+q
\put(200,35){$\leftarrow$}
\put(200,15){$l+q$}
%\put(340,90){$k+q=0$}
%
\end{picture} 

\be
  {\cal G}[\f,\y ] = g_{X} m      \int d^4 l    d^4 q d^4 k\; \d^4(k+q)
 \fr{ \f \cdot l \cdot \y } {  (l+q)^2  
 (l)^2 } 
\ee
  This looks linearly divergent.  If diagrams like this are preseent, then we would expect to need a new counterterm
 \be
\cA_{\rm New\; Counterterm} = \int d^4 x \lt \{ 
m  \;     \f \cdot\pa  \cdot \y
\rt \} 
\ee 
This term is not in the original action.  We have put a term like this into the matrix in section \ref{fermionmatrix} at the term $a_{10}  m  \pa^{\b \dot \a}$ in the matrix 
(\ref{fermionmatrixeq}).  But as a first effort we would remove that on the basis of the reasoning in section \ref{cantrans} below, and the notion that these terms do not contribute to the propagator that we want to find.  It might turn out to be needed when we do the coding to find the propagator but our first effort as set out in section \ref{conclusion} will not include this term.

Similar remarks apply to the following terms:

\subsection{The new counterterm  $  {\cal G}[\c,\lam ]\Ra \cA_{\rm New\; Counterterm} = \int d^4 x \lt \{ 
      \c \cdot\pa  \cdot \lam
\rt \} $}

%\normalsize
%template-improved March 1, 2000:
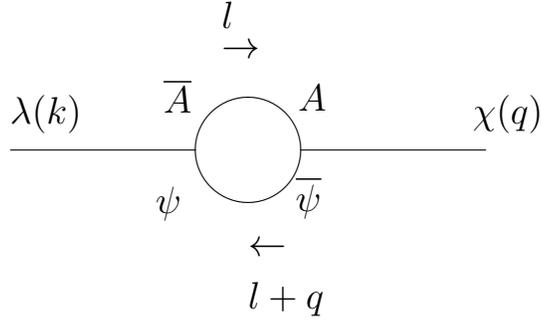
\begin{picture}(250,150)
%\put(100,02){Feynman Diagram for ${\cal G}_1[\f, L, \oy_{K}] ; k+q=0$ }
%
%
%
%\put(0,150){ $ integral $  }
%
% four point at left:
%\put(180,75){\line(-1,-1){50}}
%\put(180,75){\line(-1,1){50}}
% top oblique at right:
%\put(210,92){\line(1,2){15}}
% bottom oblique at right:
%\put(210,58){\line(1,-2){20}}
% horizontal at right:
\put(220,75){\line(1,0){70}}
% four point at right:
%\put(220,75){\line(1,1){50}}
%\put(220,75){\line(1,-1){50}}
% horizontal at left:
\put(180,75){\line(-1,0){70}}
\put(200,75){\circle{40}}
%
% outer fields
%a1
%\put(95,20){${\ov a}_1(p)$}
%a2
%\put(95,120){${\ov a}_2(p)$}
%a3
%\put(230,120){$\f^{\dot \a}(0)$}
%a4
%\put(280,120){${\ov a}_4(p)$}
%a5
\put(285,85){$\c(q)$ }
%a6
%\put(280,20){${\ov a}_6(q)$}
%a7
%\put(235,20){${\ov \lam}_{ \dot \a}(q)$}
% inner Fields:
%w
\put(110,85){{$\lam (k)$}}
%b1
\put(168,89){$\A$}
%b2
%\put(190,99){${\ov\y_K }$}
%b4
\put(219,90){$A$}
%b5
\put(218,52){$\oy$}
%b7
%TT\put(195,42){${\ov L}$}
%b8
\put(165,51){$\y $ }
% masses
%m1
%\put(186,88){$\ast$}
%\put(184,81){${m }$}
%m2
%\put(207,89){$\ast$}
%\put(199,84){${m}$}
%m3
%\put(217,73){$\ast$}
%\put(201,72){${m_3}$}
%m4
%\put(207,55){$\ast$}
%\put(200,64){${m}$}
%m5
%\put(186,56){$\ast$}
%\put(185,64){$m_5$}
%m6
%\put(177,73){$\ast$}
%\put(185,73){${m}$}
%
% momenta
%l-p
\put(190,110){$\rightarrow$}
\put(190,121){$l$}
%l
%\put(250,75){$\downarrow$}
%\put(265,75){$l$}
%l+q
\put(200,35){$\leftarrow$}
\put(200,15){$l+q$}
%\put(340,90){$k+q=0$}
%
\end{picture} 

\be
  {\cal G}[\c,\lam ] = g_{X}  g      \int d^4 l    d^4 q d^4 k\; \d^4(k+q)
 \fr{ \c \cdot(l+q)\cdot  \lam  } {  (l+q)^2  
 (l)^2 } 
\ee
  This looks linearly divergent.  If other diagrams do not cancel it, then one would expect to need a new counterterm
 \be
\cA_{\rm New\; Counterterm} = \int d^4 x \lt \{ 
      \c \cdot\pa  \cdot \lam
\rt \} 
\ee 
This term is not in the original action.

\subsection{The new counterterm  $  {\cal G}[\f,\lam ]\Ra \cA_{\rm New\; Counterterm} = \int d^4 x \lt \{ 
      \f  \Box   \ov \lam
\rt \} $}

%\normalsize
%template-improved March 1, 2000:
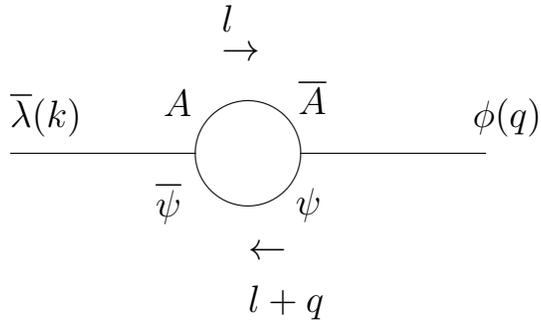
\begin{picture}(250,150)
%\put(100,02){Feynman Diagram for ${\cal G}_1[\f, L, \oy_{K}] ; k+q=0$ }
%
%
%
%\put(0,150){ $ integral $  }
%
% four point at left:
%\put(180,75){\line(-1,-1){50}}
%\put(180,75){\line(-1,1){50}}
% top oblique at right:
%\put(210,92){\line(1,2){15}}
% bottom oblique at right:
%\put(210,58){\line(1,-2){20}}
% horizontal at right:
\put(220,75){\line(1,0){70}}
% four point at right:
%\put(220,75){\line(1,1){50}}
%\put(220,75){\line(1,-1){50}}
% horizontal at left:
\put(180,75){\line(-1,0){70}}
\put(200,75){\circle{40}}
%
% outer fields
%a1
%\put(95,20){${\ov a}_1(p)$}
%a2
%\put(95,120){${\ov a}_2(p)$}
%a3
%\put(230,120){$\f^{\dot \a}(0)$}
%a4
%\put(280,120){${\ov a}_4(p)$}
%a5
\put(285,85){$\f(q)$ }
%a6
%\put(280,20){${\ov a}_6(q)$}
%a7
%\put(235,20){${\ov \lam}_{ \dot \a}(q)$}
% inner Fields:
%w
\put(110,85){{$\ov \lam(k)$}}
%b1
\put(168,89){$A$}
%b2
%\put(190,99){${\ov\y}$}
%b4
\put(219,90){$\A$}
%b5
\put(218,52){$\y$}
%b7
%TT\put(195,42){${\ov L}$}
%b8
\put(165,51){$\oy $ }
% masses
%m1
%\put(186,88){$\ast$}
%\put(184,81){${m }$}
%m2
%\put(207,89){$\ast$}
%\put(199,84){${m}$}
%m3
%\put(217,73){$\ast$}
%\put(201,72){${m_3}$}
%m4
%\put(207,55){$\ast$}
%\put(200,64){${m}$}
%m5
%\put(186,56){$\ast$}
%\put(185,64){$m_5$}
%m6
%\put(177,73){$\ast$}
%\put(185,73){${m}$}
%
% momenta
%l-p
\put(190,110){$\rightarrow$}
\put(190,121){$l$}
%l
%\put(250,75){$\downarrow$}
%\put(265,75){$l$}
%l+q
\put(200,35){$\leftarrow$}
\put(200,15){$l+q$}
%\put(340,90){$k+q=0$}
%
\end{picture} 

\be
  {\cal G}[\c,\lam ] = g_{X}  g      \int d^4 l    d^4 q d^4 k\; \d^4(k+q)
 \fr{    \f  \cdot  l \cdot (l+q) \cdot \ov \lam } {  (l+q)^2  
 (l)^2 } 
\ee
  This looks quadratically divergent.  If other diagrams do not cancel it, then one would expect to need a new counterterm
 \be
\cA_{\rm New\; Counterterm} = \int d^4 x \lt \{ 
      \f   \Box \ov \lam^0
\rt \} 
\ee 
This term is not in the original action.

We will not try to look at the possible new terms for the bosons in this paper, but it seems likely that there are some.  These diagrams might not exist, and no care to write out the entire action, in terms of the new fields, has been taken here.  If care is needed, that will be done in a future set for coding for the more difficult problem that might emerge if the first attempt is unsuccessful.

   \section{The \CaT s for the ZX Sector}

\la{cantrans}

   \subsection{Generating Functional for the new fermionic quadratic terms from section \ref{diagrams} in the ZX sector}
\la{CaTsfortheZsector}

The business of finding a canonical transformation can be found in \ci{ Kluberg-Stern:1974iel,  Kluberg-Stern:1975ebk, Kluberg-Stern:1974nmx,Deans:1978wn,Dixon:1974ss,Dixon:1975si}. Our notation follows that in  \ci{Dixon:1975si}.  The background field gauge is particulary diffficult when there are \ei s present.

  The following is a generator for the canonical transformation for the spinors:
\be
\cG [\rm   Fermi] = \int d^4 x \eb
\lt \{ 
 L^{\a} (\b_{1} \lam'_{\a}+i \b_{2} \ov\c'_{\a})   
+  K^{\dot \a} (\b_{3} \c'_{\dot \a}+i\b_{4} \ov\lam'_{\dot \a})   
\ebp + G^{\dot \a} (\b_{5}\f'_{\dot \a}) 
+Y^{\a}(\b_{6}\y'_{\a}+ \b_{7} g m \ov \f'_{\a})
\rt \} + *
\ee
where we put $gm$ in the last term $ \b_{7} g m \ov \f'_{\a}$ for future convenience.
and so for example we have
\be
\fr{\d  \cG [\rm   Fermi ] }{\d L^{\a}}
= \lam_{\a} = 
(\b_{1} \lam'_{\a}+i \b_{2} \ov\c'_{\a})  
\ee
\be
\fr{\d  \cG [\rm   Fermi ] }{\d K^{\dot \a}}
= \c_{\dot \a} =
 (\b_{3} \c'_{\dot\a}+i \b_{4} \ov\lam'_{\dot \a}) 
\ee
\be
\fr{\d  \cG [\rm   Fermi ] }{\d G^{\dot \a}}
= \f_{\dot\a} = 
\b_{5}  \f'_{\dot\a}
\ee
\be
\fr{\d  \cG [\rm   Fermi ] }{\d Y^{\a}}
= \y_{\a} =
 (\b_{6} \y'_{\a}+ \b_{7} g m \ov\f'_{\a})  
\ee

\be
\fr{\d  \cG [\rm   Fermi ] }{\d \ov L^{\dot\a}}
=\ov \lam_{\dot\a} =
 (\b_{1}\ov \lam'_{\dot\a}-i \b_{2}  \c'_{\dot \a}) 
\ee
\be
\fr{\d  \cG [\rm   Fermi ] }{\d \ov K^{\a}}
= \ov\c_{\a} =
 (\b_{3} \ov\c'_{\a}-i \b_{4} \lam'_{\a}) 
\ee
\be
\fr{\d  \cG [\rm   Fermi ] }{\d \ov G^{  \a}}
= \ov\f_{ \a} = \b_{5} \ov \f'_{\a}
\ee
\be
\fr{\d  \cG [\rm   Fermi ] }{\d \ov Y^{\dot\a}}
= \oy_{\dot\a} =
 (\b_{6} \oy'_{\dot\a}+ \b_{7} g m \f'_{\dot \a})  
\ee
There are also corresponding transformations of the pseudofields of course:
\be\fr{\d  \cG [\rm   Fermi ] }{\d \ov\lam'_{ \dot \a}} 
=\ov L^{` \dot \a} = (\b_{1} \ov L^{\dot \a}- i \b_{4}  K^{\dot \a})    
\ee

   \subsection{Generating Functional for the bosonic new terms in the ZX sector}

  The following are generators for the canonical transformation for the bosons, although we did not try to write down any counterterms for bosons in section  \ref{diagrams}:
\be
\cG_{\rm Bosons\;Real}=
\int d^4 x \lt \{
\S^{\a \dot \b}\lt (
\g_1  U'^{\a \dot \b} + \g_2  \cJ'^{\a \dot \b} 
\rt )
\ebp
+
\W^{\a \dot \b}\lt (
\g_3  \cJ'^{\a \dot \b} 
  + \g_4 U'^{\a \dot \b}
\rt )
+
\X^{\a \dot \b}\lt (
\g_5  \cR'^{\a \dot \b} 
\rt )
\ebp
+
\D \lt (
\g_6  D' + 
\g_7  m ( A' + \A' )\rt )
+
\ebp
+
\z \lt (
\g_8 Z'
\rt )
+
H \lt (
\g_9 \h'
\rt )
+
W \lt (
\g_{10} \w'
\rt )
\rt \}
\ee

\be
\cG_{\rm Bosons\;Complex}=
\int d^4  x   \lt \{  \G \lt ( \g_{11}
 A   \rt )
\ebp
+
\Lam \lt (
\g_{12}  F
 + i \g_{13}
(  A'  - \A')
 \rt )
\rt \}
\ee
with 
\be
\fr{\d \cG_{\rm Bosons }}{\d \S^{\a \dot \b}}=
U_{\a \dot \b}=
\lt (
\g_1  U'_{\a \dot \b} + \g_2  \cJ'_{\a \dot \b} 
\rt )
\ee

\be
\fr{\d \cG_{\rm Bosons }}{\d \W^{\a \dot \b}}=
U_{\a \dot \b}=
\lt (
\g_3  \cJ'_{\a \dot \b} + \g_4  U'_{\a \dot \b}  
\rt )
\ee

\be
\fr{\d \cG_{\rm Bosons }}{\d \X^{\a \dot \b}}=
\cR_{\a \dot \b}=
\lt (
\g_5  \cR'_{\a \dot \b} 
\rt )
\ee

\be
\fr{\d \cG_{\rm Bosons }}{\d \D}=
D=
 \lt (
\g_6  D' + 
\g_7  m ( A' + \A' )\rt )
\ee

\be
\fr{\d \cG_{\rm Bosons }}{\d \Lam}=
F=
 \lt (
\g_{12}  F
 + i \g_{13}
(  A'  - \A')
 \rt )
\ee

It is not yet clear whether these are needed here to solve the tachyon problem. We hope not, but it seems conceivable.  

  \section{ Coding Problems to be Solved}
\la{coding}

The XM is made from the SSM, which is already a  formidably complicated model.  There are a number of computer programs designed to extract experimental results from that theory and its variations \ci{SARAH}.  They are not designed for the present issues, but no doubt there are many common threads. The present set of problems needs to concentrate on the BRS nilpotence and also on the new kinds of SUSY representations such as the CDSS and the problem of finding tachyon-free solutions of the coefficients.  

One of the important equations to be solved in this manner is the equation derived like the one in section \ref{tachyon}, that says that there are no tachyons.  This imposes constraints on the coefficients.  It is not at all clear that the theory can bear this kind of constraint, and a careful coding solution is needed.

In this paper we show the form of various expressions, but we are not trying to develop the theory in terms of the coefficients in E6.  We utilize the concept of modularity here, hoping to isolate errors in small parts.  Ultimately, of course, if the theory actually works, we will want to close the gaps.

  \section{Some Remarks about the 
Spectral Sequence}

\la{specseq}

  It took fifty years to find  the  \ei s.  The most obvious reason for that long time is that the  \ei s are very well hidden.  The spectral sequence seems to be necessary to find them, and it is notorious for its obscurity,  and for its power \ci{mcleary}.  The crucial progress was to note that  SUSY, like the Lie groups,   has a BRS cohomology that is governed by its structure constants. The choice of the grading for the spectral sequence in \ci{E2} uses that fact. Also, the derivation of the form of the \ei s from the form of the space $E_{\infty}$ is really very far from obvious, as can be seen from \ci{E2}. The constraint arises from what is usually called the Elizabethan drama part of the spectral sequence.  For example the derivation of the constants in equation (\ref{valuesofbs}) is needed to implement the mapping $ E_{\infty} \ra \cH$, as shown in \ci{E4}. Solving the constraint is non-trivial so it is noteworthy that a solution exists in the case treated in E6.  Some   useful  texts on supersymmetry, including chiral superfields,  are  \ci{haberetal,xerxes,bagger,weinberg}. Recent reviews about the SSM and SUSY experiments are
\ci{Allanach:2024suz,D’Onofrio}.

It seems likely that we should continue to examine the spectral sequence, by adding more complicated SUSY representations, in addition to the CDSS.   After all, the superstring \ci{witten} uses SUSY representations of all spins.  Should we expect that there are \ei s of all spins?  For example, can the work in \ci{E2} be expanded to include objects like those in  \ci{dixminram}? Would that be useful?

\section{The error from the general integral over superspace}
\la{badintegral}
  The  existence of  \ei s is inconsistent with the  results of a paper published in 1980  \ci{pigsib}. That paper was a very general proof that the BRS cohomology of SUSY in 3+1 dimensions must be trivial.     It  is often cited still, for example in the recent papers  \ci{Ferrero:2023xsf,imbimbo,minasian}.  But the paper \ci{pigsib} is incorrect.  Since the results in the present paper contradict the result in \ci{pigsib}, it is necessary to   explain what is wrong with that paper. The mistake in that paper is very simple. Its very general argument is based on the superspace Grassmann variable $\q_{\a}$ and its complex conjugate $\oq_{\dot \a}$.  These are treated  as though they were free variables, and as though all relevant integrals are integrated over all four of these variables.  However, this is not true for chiral or antichiral integrals, which leave off either  $\oq_{\dot \a}$ or $\q_{\a}$  in the integrations, respectively \ci{bagger}.   But \ci{pigsib} does not discuss chiral superfields. I am grateful to the first author of \ci{pigsib} for a useful discussion of this problem in 1990, after the paper \ci{holes} was written. As I recall, he noted the problem raised here, during that discussion.   For example, the important equation (24) of   \ci{pigsib} is clearly a general superfield.
Up until now it has seemed to this author that this error was not very important, since it seemed that any non-trivial cohomology in SUSY theories was not Lorentz invariant.  But now with the present results the error needs to be corrected, since the exotic invariants are indeed Lorentz invariant and may even be important. In addition, it seems that, with the remarks in the previous section \ref{specseq}, there may be very many more examples like the presently known \ei s.

  \section{Discussion of the Canonical Transformations and Tachyons in the ZX Sector}

\la{discuss}

It is clear that there are a number of complicated problems that are already solved in this set of papers.  
Are the solutions correct?  For example, the calculation of (\ref{valuesofbs}) and of section \ref{solforquad} should be verified by a computer program.   

From the theory point of view, it is not clear (to this author) what  relation exists between the elimination of tachyons using results like those in section \ref{tachyon} and the generation of canonical transformations as in section \ref{cantrans}.  Normally one expects that the canonical transformations  can be reversed, so that one can deal with the simple action before the canonical transformation for many purposes. But there is a new problem here too.  In order to get some of the diagrams that give rise to the new counterterms that generate the canonical transformations, we need to assume that we know the propagators, but those are what we are trying to find.  Is this some sort of circular problem?  Surely not?  
Also the  tachyon constraints that arise, like those in section \ref{tachyon}, confuse the matter even further.  A computer program might help to see these questions more clearly.

  \section{Conclusion}
\la{conclusion}

The discovery of the exotic invariants raises some new possibilities for supersymmetry.  Probably the most promising feature is that the XM seems to contain a mechanism for mass splitting that seems well adapted to give the experimental result that there are suppressed neutral currents, as noted in E6.  The XM also solves some of the problems that are present for spontaneous and explicit breaking of \susy.
But it is far from clear whether the XM is a healthy model. 

To determine that, the first  calculation should be to check the action in section \ref{solforquad}, to ensure that it is consistent with sections \ref{Pseudoquad} and \ref{MEquad}.  Then that solution should be used for the forms in sections \ref{fermionmatrix} and \ref{bosonmatrix}.  Those forms should then be solved for the propagators, while imposing the constraints that the theory be free of tachyons, as suggested in   section \ref{tachyon}. At first, we should assume that there are no  new counterterms as suggested in section \ref{diagrams}. This set of  steps  require a computer program.  If that program is successful, then new counterterms like those in section \ref{diagrams} can presumably be removed with canonical transformations like those in section \ref{cantrans}.   It is conceivable however, that a more complicated solution exists.  

The main questions that relate to these steps are: Do solutions for the tachyon problem in the ZX sector exist?  Can the higher derivatives be handled sensibly, at least at low orders? Do the canonical transformations get generated partly by the propagators we are trying to find using them? Because the ZX sector is so complicated, computer programs are needed to determine these questions.  If the theory makes sense at this level, it might then be possible to calculate the one loop SUSY violating corrections to the  mass spectra in the quark, lepton, charged vector, photon and Higgs sectors. At some point this Exotic Model, if it works, might lead to experimental predictions.  That would still leave all sorts of issues of course, including the non-renormalizability of this model and its connection to supergravity and the superstring.

\begin{center}
 { Acknowledgments}
\end{center}
\vspace{.1cm}

  I thank  Carlo Becchi,  Margaret Blair, Friedemann Brandt, Philip Candelas, David Cornwell,   James Dodd, Mike Duff, Sergio Ferrara, Richard Golding, Dylan Harries, Marc Henneaux, Chris T.  Hill,  D.R.T. Jones, Olivier Piguet, Antoine van Proeyen,  Pierre Ramond,   Peter Scharbach,      Mahdi Shamsei, Kelly Stelle, Sean Stotyn, Xerxes Tata, J.C. Taylor,  Peter West and Ed Witten for stimulating correspondence and conversations.   I also express appreciation for help in the past from William Deans, Lochlainn O'Raifeartaigh, Graham Ross, Raymond Stora, Steven Weinberg, Julius Wess and Bruno Zumino. They are not replaceable and they are missed.  I   thank  Doug Baxter,  Margaret Blair,  Murray Campbell, David Cornwell, James Dodd, Davide Rovere,   Pierre Ramond, Peter Scharbach and Mahdi Shamsei for recent, and helpful, encouragement to carry on with this work. I also express appreciation to Dylan Harries and  to Will, Dave and Peter Dixon and Vanessa McAdam for encouraging and teaching me to use coding.

%\tableofcontents

 \tiny 
\articlenumber\\
Aug 7, 2025

\end{document}

%% file: Johnmacros29.tex
 \newcommand{\susy}{{supersymmetry}}

\newcommand{\Pf}{{PseudoField}}

  \newcommand{\cdss}{chiral dotted spinor superfield}

\newcommand{\XM}{{Exotic Model}}

\newcommand{\CaT}{Canonical Transformation}

\newcommand{\ME}{Master Equation}

\newcommand{\cJ}{{\cal J}}

\newcommand{\cM}{{\cal M}}

\newcommand{\CDSS}{{Chiral Dotted Spinor Superfield}}

\newcommand{\cG}{{\cal G}}

\newcommand{\cA}{{\cal A}}

\newcommand{\CC}{Complex Conjugate}

\newcommand{\Exists}{\bm{\exists}\kern-0.6em\bm{\exists}}

\newcommand{\EI}{Exotic Invariant}
\newcommand{\ei}{exotic invariant}

\newcommand{\SSM}{Supersymmetric Standard Model}

\newcommand{\SM}{Standard Model}

\newcommand{\sg}{$\rm SU(3) \times SU(2) \times U(1)$}

\newcommand{\bsg}{$\rm SU(3)  \times U(1)$}

\newcommand{\cR}{{\cal R}}

\newcommand{\cL}{{\cal L}}
\newcommand{\bt}{\begin{tabular}{c}}
\newcommand{\et}{\end{tabular}}

\newcommand{\eb}{\ee\be } 
\newcommand{\ebp}{\rt.\ee\be\lt.}

\newcommand{\bmat}{\lt ( \begin{array} }
\newcommand{\emat}{  \end{array} \rt )}

\newcommand{\cH}{{\cal H}}

\newcommand{\oX}{{\ov X}}

\newcommand{\oq}{{\ov \q}}

\newcommand{\oK}{{\ov K}}

\newcommand{\oY}{{\ov Y}}

\newcommand{\oy}{{\ov \y}}

\newcommand{\ove}{{\ov e}}

\newcommand{\oC}{{\ov C}}
\newcommand{\oL}{{\ov L}}

\newcommand{\oF}{{\ov F}}
\newcommand{\A}{{\ov A}}

\renewcommand{\a}{\alpha}	
\renewcommand{\b}{\beta}
\newcommand{\g}{\gamma}
\renewcommand{\d}{\delta}
\newcommand{\e}{\epsilon}
\newcommand{\ve}{\varepsilon}
\newcommand{\z}{\zeta}
\newcommand{\h}{\eta}
\newcommand{\q}{\theta}

\newcommand{\lam}{\lambda}
\newcommand{\m}{\mu}
\newcommand{\n}{\nu}	
\newcommand{\x}{\xi}

\newcommand{\f}{\phi}
\renewcommand{\c}{\chi}
\newcommand{\y}{\psi}
\newcommand{\w}{\omega}
\newcommand{\G}{\Gamma}
\newcommand{\D}{\Delta}
	
\newcommand{\Lam}{\Lambda}
\newcommand{\X}{\Xi}
	
\renewcommand{\S}{\Sigma}

\newcommand{\W}{\Omega}
\newcommand{\la}{\label}
\newcommand{\ci}{\cite}

\newcommand{\ds}{\documentstyle}	
\newcommand{\fr}{\frac}

\newcommand{\pa}{\partial}
\newcommand{\ov}{\overline}
\newcommand{\br}{\begin{rant}}
\newcommand{\er}{\end{rant}}
\newcommand{\beC}{\begin{Conjecture}}
\newcommand{\eeC}{\end{Conjecture}}
\newcommand{\be}{\begin{equation}}
\newcommand{\ee}{\end{equation}}
\newcommand{\ba}{\begin{array}} 
\newcommand{\ea}{\end{array}}
\newcommand{\bea}{\begin{eqnarray}}
\newcommand{\eea}{\end{eqnarray}}
\newcommand{\ra}{\rightarrow}
\newcommand{\Ra}{\Rightarrow}

\newcommand{\lt}{\left}
\newcommand{\rt}{\right}

\newcommand{\ben}{\begin{enumerate}}
\newcommand{\een}{\end{enumerate}}
\newcommand{\bitem}{\begin{itemize}}
\newcommand{\eitem}{\end{itemize}}

%% file: E7Aug7-8Final.bbl
\begin{thebibliography}{99}



\bibitem{SARAH} SARAH is a mathematica program for supersymmetric theories:  https://sarah.hepforge.org/



 \vspace{1cm}
Recent reviews about supersymmetry and experiment

\bibitem{Allanach:2024suz}
B.~Allanach and H.~E.~Haber,
``Supersymmetry, Part I (Theory),'' (Particle Data Group)
[arXiv:2401.03827 [hep-ph]].
%6 citations counted in INSPIRE as of 25 Apr 2025

\bibitem{D’Onofrio} M. D’Onofrio  and F. Moortgat,  
 ``Supersymmetry, Part II (Experiment)'' (Particle Data Group)
Revised August 2023  



 \vspace{1cm}
A few useful textbooks

\bibitem{haberetal}  Dreiner, Herbi K., Howard E. Haber, and Stephen P. Martin, ``From Spinors to Supersymmetry'' (Cambridge: Cambridge University Press, 2023) 
 
\bibitem{xerxes} 
  H.~Baer and X.~Tata,
  ``Weak scale supersymmetry: From superfields to scattering events'',
  Cambridge, UK: Univ. Pr. (2006)  
  
    

\bibitem{bagger} 
J.~Wess and J.~Bagger,
``Supersymmetry and supergravity,''
Princeton University Press, 1992,
ISBN 978-0-691-02530-8
 
 

 \bibitem{weinberg} Steven Weinberg: ``The Quantum Theory of fields" Volume 3, Cambridge University Press, ISBN 052155002.  
 \bibitem{quarkmasses}Steven Weinberg, ``The quantum theory of fields volume 2, Cambridge 2005,  
  
   
\bibitem{witten}  
M.~B.~Green, J.~H.~Schwarz and E.~Witten,
``Superstring Theory Vol. 1: 25th Anniversary Edition,''
Cambridge University Press, 2012,
ISBN 978-1-139-53477-2, 978-1-107-02911-8
 
       
\bibitem{mcleary} J.~ McCleary, ``A User's Guide to Spectral Sequences", Second Edition,
Cambridge 2001.  
 







  


\vspace{1cm}
Some Papers about new counterterms and canonical transformations to generate them:


%\cite{Kluberg-Stern:1974iel}
\bibitem{Kluberg-Stern:1974iel}
H.~Kluberg-Stern and J.~B.~Zuber,
``Ward Identities and Some Clues to the Renormalization of Gauge Invariant Operators,''
Phys. Rev. D \textbf{12}, 467-481 (1975)
doi:10.1103/PhysRevD.12.467
%308 citations counted in INSPIRE as of 05 Aug 2025

%\cite{Kluberg-Stern:1975ebk}
\bibitem{Kluberg-Stern:1975ebk}
H.~Kluberg-Stern and J.~B.~Zuber,
``Renormalization of Nonabelian Gauge Theories in a Background Field Gauge. 2. Gauge Invariant Operators,''
Phys. Rev. D \textbf{12}, 3159-3180 (1975)
doi:10.1103/PhysRevD.12.3159
%281 citations counted in INSPIRE as of 05 Aug 202


%\cite{Kluberg-Stern:1974iel}
%\cite{Kluberg-Stern:1975ebk}%\cite{Kluberg-Stern:1974nmx}
\bibitem{Kluberg-Stern:1974nmx}
H.~Kluberg-Stern and J.~B.~Zuber,
``Renormalization of Nonabelian Gauge Theories in a Background Field Gauge. 1. Green Functions,''
Phys. Rev. D \textbf{12}, 482-488 (1975)
doi:10.1103/PhysRevD.12.482
%348 citations counted in INSPIRE as of 05 Aug 2025





\bibitem{Dixon:1974ss}
J.~A.~Dixon and J.~C.~Taylor,
``Renormalization of Wilson operators in gauge theories,''
Nucl. Phys. B \textbf{78}, 552-560 (1974)
doi:10.1016/0550-3213(74)90598-7
%100 citations counted in INSPIRE as of 05 Aug 2025

 %\cite{Kluberg-Stern:1974iel}
%\cite{Kluberg-Stern:1975ebk}%\cite{Kluberg-Stern:1974nmx}
%\bibitem{Deans:1978wn}
\bibitem{Dixon:1975si}
J.~A.~Dixon,
``Field Redefinition and Renormalization in Gauge Theories,''
Nucl. Phys. B \textbf{99}, 420-424 (1975)
doi:10.1016/S0550-3213(75)80018-6
%44 citations counted in INSPIRE as of 05 Aug 2025
 
 %\cite{Kluberg-Stern:1974iel}
%\cite{Kluberg-Stern:1975ebk}%\cite{Kluberg-Stern:1974nmx}
\bibitem{Deans:1978wn}
W.~S.~Deans and J.~A.~Dixon,
``Theory of Gauge Invariant Operators: Their Renormalization and S Matrix Elements,''
Phys. Rev. D \textbf{18}, 1113-1126 (1978)
doi:10.1103/PhysRevD.18.1113
%59 citations counted in INSPIRE as of 05 Aug 2025

\vspace{1cm}
An important and general but incorrect  paper (see section (\ref{badintegral}):  


\bibitem{pigsib}
%\cite{Piguet:1980fa}
%\bibitem{Piguet:1980fa}
O.~Piguet, K.~Sibold and M.~Schweda,
``GENERAL SOLUTION OF THE SUPERSYMMETRY CONSISTENCY CONDITIONS,''
Nucl. Phys. B \textbf{174}, 183-188 (1980)
doi:10.1016/0550-3213(80)90197-2
%55 citations counted in INSPIRE as of 27 Nov 2024

\vspace{1cm}
There are many papers that  that refer to the above wrong paper \ci{pigsib} as though it were correct. Here are the three most recent ones:


%\cite{Ferrero:2023xsf}
\bibitem{Ferrero:2023xsf}
R.~Ferrero, M.~B.~Fr{\"o}b and W.~C.~C.~Lima,
%``Heat kernel coefficients for massive gravity,''
J. Math. Phys. \textbf{65}, no.8, 082301 (2024)
doi:10.1063/5.0196609
[arXiv:2312.10816 [hep-th]].
%9 citations counted in INSPIRE as of 06 Aug 2025


\bibitem{imbimbo}
%\cite{Frob:2021sao}
M.~B.~Fr\"ob, C.~Imbimbo and N.~Risso,
``Deformations of supergravity and supersymmetry anomalies,''
JHEP \textbf{12}, 009 (2021)
doi:10.1007/JHEP12(2021)009
[arXiv:2107.03401 [hep-th]].
%4 citations counted in INSPIRE as of 27 Nov 2024




\bibitem{minasian}
R.~Minasian, I.~Papadimitriou and P.~Yi,
``Anomalies and supersymmetry,''
Phys. Rev. D \textbf{105}, no.6, 065005 (2022)
doi:10.1103/PhysRevD.105.065005
[arXiv:2104.13391 [hep-th]].
%4 citations counted in INSPIRE as of 27 Nov 2024

\vspace{1cm}

{ The E Series: }The following 8 papers, about exotic invariants,  form a series  starting at  \ci{E1}, which is E1.  They are labelled (En), where $n=1,2\cdots$.   



\bibitem{E1}   
J.~A.~Dixon,  `` Supersymmetry anomalies, exotic pairs and the  supersymmetric standard mode (E1)",   [arXiv:2407.13673]. 


 \bibitem{E2}     Ibid. 
``The BRS Cohomology of the  Wess Zumino Chiral Scalar supersymmetric model with exotic pairs and exotic triplets (E2)'', 
[arXiv:2507.14174]



 \bibitem{E3}     Ibid. 
  ``The  simplest Exotic Invariant in the BRS cohomology of 
  supersymmetry in 3+1 dimensions (E3)''
Preprint to be issued soon.

\bibitem{E4}Ibid. 
  ``The EP Model with Completion Terms (E4)'' Preprint to be issued soon.
  
    \bibitem{E5}Ibid. ``The EP Model with U(1) (E5)'',  Preprint to be issued soon.



    \bibitem{E6}Ibid. ``The Supersymmetric Standard Model, combined with a special \EI, yields a new kind of SUSY mass splitting (E6)."  [arXiv:2507.14381].


  



    \bibitem{E7}Ibid. ``The Free Massive Quadratic Action from the \XM.   (E7)''    This is the present Preprint,  issued in August 2025.



\bibitem{E??}Ibid. 
  ``The BRS cohomology of SUSY Including Gauge Theory and the \CDSS\ (E??)''. Preprint to be issued, but not soon.


\vspace{1cm}
Some early work on BRS cohomology of SUSY by the author and collaborators:


\bibitem{holes} 
%\cite{Dixon:1990jv}
J.~A.~Dixon,
``Supersymmetry is full of holes'',
Class. Quant. Grav. \textbf{7}, 1511-1521 (1990)
doi:10.1088/0264-9381/7/8/026



\bibitem{holescommun} 
Ibid. ,
``BRS cohomology of the chiral superfield'',
Commun. Math. Phys. \textbf{140}, 169-201 (1991)


\bibitem{dixminram}  
J.~A.~Dixon, R.~Minasian and J.~Rahmfeld,
"Higher spin BRS cohomology of supersymmetric chiral matter in D = 4'',
Commun. Math. Phys. \textbf{171}, 459-474 (1995)
[arXiv:hep-th/9308013 [hep-th]].


 \end{thebibliography}
